\documentclass[a4paper,11pt]{article}
\usepackage{jheppub} % for details on the use of the package, please see the JINST-author-manual

\usepackage{lineno}%标行号需要
\usepackage{braket}%狄拉克括号需要
\usepackage{booktabs}%精细表格需要
\usepackage{xcolor}%标记颜色需要

\usepackage{booktabs, threeparttable}
\usepackage{tabularx}
\usepackage{array}
\newcolumntype{Y}{>{\raggedright\arraybackslash}X}

\arxivnumber{} % if you have one

\title{\boldmath Long-range multipartite entanglement in holographic gapless systems}

\author[a]{Xiantong Chen,}
\author[a,b]{Xuanting Ji,}
\author[c]{Xin-Xiang Ju,}
\author[b]{Wen-Peng Li,}
\author[a]{Ya-Wen Sun}
\affiliation[a]{School of Physical Sciences,\\ 
    University of Chinese Academy of Sciences, Beijing 100049, China
}
\affiliation[b]{Department of Applied Physics, College of Science, \\
    China Agricultural University, Beijing 100083, China
}
\affiliation[c]{
    Institute for Advanced Study,\\
    Tsinghua University, Beijing 100084, China
}
\emailAdd{{chenxiantong23@mails.ucas.ac.cn}, {jixuanting@cau.edu.cn}, {juxinxiang21@mails.ucas.ac.cn}, {liwenpengsci@cau.edu.cn}, {yawen.sun@ucas.ac.cn}}

\abstract{Gapless quantum systems support correlations over arbitrarily long distances, giving rise to power-law long-range entanglement. We investigate long-range entanglement at strong coupling for three-dimensional gapless systems using holography, asking whether multipartite entanglement can exhibit scale-growing behavior and become enhanced at large distances. We show that a broad class of multipartite entanglement quantities share the same leading large-distance scaling exponent determined by the IR geometry. To realize  different scaling regimes, we consider hyperscaling-violating IR geometries. Depending on the parameters, long-distance multipartite entanglement can decay, become logarithmic, grow subextensively, or reach a volume law. We also analyze an anisotropic IR geometry \(\mathrm{AdS}_{3}\times\mathbb{R}^{2}\), where long-range multipartite entanglement survives along one direction but becomes short-ranged in the transverse gapped directions. These results show that holographic gapless phases can support rich and enhanced long-range multipartite entanglement, providing a nonlocal characterization of the underlying IR physics.}

\begin{document}
\raggedbottom
\setlength{\parskip}{0pt}
\setlength{\parindent}{2em}
\maketitle
\flushbottom

%%%%%%%%%%%%%%%%%%%%%%%%%不想自动分页include改用input
\section{Introduction}
    \label{Section1}
    Quantum entanglement is one of the most fundamental concepts in modern physics, serving as a key resource in quantum information theory and providing deep insights into the nature of quantum many-body systems. In strongly correlated systems, entanglement entropy has become a powerful tool for characterizing quantum phases, quantum criticality, and topological order \cite{Levin2006,Wen2013,Kitaev2006,Mazza2026}. The behavior of how the size of the subsystems affects the entanglement quantities provides a universal fingerprint that distinguishes different phases of matter: gapped phases obey an area law, while gapless critical systems exhibit logarithmic or power-law scaling behavior at large distance scales. 

    Entanglement entropy, however, characterizes correlations only across a single bipartition. Gapless states contain correlations over arbitrarily long length scales, and these correlations can connect several spatial regions in an intrinsically multipartite manner. Understanding their IR structure therefore requires studying not only whether long-distance entanglement survives, but also how multipartite entanglement is distributed among widely separated subsystems and how it scales with the subsystem sizes. This question is particularly difficult in strongly coupled gapless systems, which often lack a quasiparticle description or a controlled perturbative low-energy theory.  Holography \cite{Witten1998,Gubser1998,Maldacena1999} provides a nonperturbative framework for addressing this problem, allowing the long-distance multipartite entanglement structure to be related systematically to the dual IR geometry.

   In holographic systems, the Ryu-Takayanagi (RT) formula provides a remarkable geometric dual of entanglement entropy \cite{Ryu2006}: the entanglement entropy of a boundary region \(A\) is given by the area of the minimal surface in the bulk that is homologous to \(A\). This profound connection between quantum information and spacetime geometry has opened new avenues for exploring the emergence of bulk spacetime from boundary quantum entanglement. In recent years, a variety of multipartite entanglement quantities have been constructed within the holographic framework, including the sign-adjusted \(n\)-partite information \((-1)^n I_n\) evaluated on a special class of configurations \cite{ju2024holographicmultipartiteentanglementupper}; the quantity \(\kappa\) , constructed from the three-body multi-entropy by subtracting its bipartite entropy contributions \cite{Harper2024, https://doi.org/10.48550/arxiv.2604.10583, https://doi.org/10.48550/arxiv.2606.04470}; the Markov gap, whose holographic evaluation involves the entanglement wedge cross section \cite{Dutta2021, Umemoto2018, Hayden2021, Zou2021, Akers_2020}; and various multipartite generalizations of the entanglement wedge cross section \cite{Bao2019, Bao2019Another}. These quantities, typically defined in terms of networks of minimal surfaces, serve as probes of different aspects of multipartite entanglement.

    In generic quantum many-body systems, multipartite entanglement is not characterized by a single universal quantity: inequivalent entanglement structures, exemplified by GHZ- and W-type states as well as more general multipartite tensor structures, can be weighted differently by different entanglement measures \cite{Bao2026}. For example, the Markov gap detects W-type tripartite entanglement but vanishes for generalized GHZ-type states, despite the latter being genuinely tripartite entangled \cite{Zou2021,https://doi.org/10.48550/arxiv.2411.11961}. Different multipartite quantities may therefore display distinct dependence on system parameters and provide complementary probes of changes in the underlying entanglement structure. Against this general background, in this work, we establish a scaling rigidity for holographic strip configurations in gapless systems: any nonvanishing multipartite quantity constructed from a finite combination of minimal surfaces has the same leading large-distance power-law exponent, which is the same exponent from the IR part of the entanglement entropy, provided that the leading contribution does not cancel. For holographic gapless states, we obtain a general formula showing that the leading IR exponent is determined by the bulk IR geometry rather than by the particular multipartite quantity. Different quantities can still differ in their coefficients, transition points, subleading behavior, and possible cancellations, but they cannot carry independently tunable leading IR exponents.

    Having established that nonvanishing RT-surface-defined multipartite quantities inherit the leading IR exponent of the IR geometry in holographic gapless systems, we can use the general formula to classify the long-distance multipartite entanglement supported by different IR geometries, especially focusing on whether enhanced long-range entanglement can be found. We first utilize  hyperscaling-violating (HV) geometries \cite{Charmousis2010,Goutraux2011,Huijse2012} as IR geometries, since they provide a particularly flexible setting in which the IR scaling can be varied continuously. Their low-energy behavior is characterized by the effective spatial dimension $d_{\mathrm{eff}}=d-\theta$, which replaces the microscopic spatial dimension in the scaling of many IR observables \cite{Fisher1986}. For example, the thermal entropy density scales as $s\sim T^{d_{\mathrm{eff}}/z_t}$, where \(z_t\) is the dynamical critical exponent \cite{Fisher1989}, while the finite strip entanglement contribution scales as $ S(l)\sim l^{1-d_{\mathrm{eff}}}$. By our general result, the same leading power governs all nonvanishing multipartite quantities constructed from a finite RT network, unless the leading term cancels.

    The resulting multipartite entanglement can therefore be tuned continuously by changing \(d_{\mathrm{eff}}\). At \(d_{\mathrm{eff}}=1\), the system exhibits the logarithmic scaling characteristic of an effectively one-dimensional critical entanglement structure. For \(0<d_{\mathrm{eff}}<1\), the long-distance multipartite contribution grows with the subsystem size while remainingsubextensive. At the limiting value \(d_{\mathrm{eff}}=0\), it becomes proportional to the strip volume, producing a volume-law multipartite contribution. This extensive ground-state entanglement has the same spatial scaling as thermal entropy, but its origin is the IR structure of the zero-temperature state. HV geometries thus provide a natural starting point for constructing IR phases with continuously tunable long-range multipartite entanglement.

    We further investigate anisotropic HV geometries in the IR, taking \(AdS_3 \times \mathbb{R}^2\) as a concrete example. This geometry can be realized by turning on a magnetic field in the \(xy\)-plane, which opens a gap in the transverse directions while preserving a gapless channel along the \(z\)-direction. The multipartite entanglement quantities in the \(z\)-direction follow a logarithmic law characteristic of a CFT\(_2\). In contrast, in the \(xy\)-plane, as the strip width increases, the RT surface undergoes a phase transition from a connected arch to two disconnected walls, leading to a pure area law. 

    The origin of enhanced long-range entanglement from these IR geometries lies in the effective dimensional reduction. The IR \(d_{\mathrm{eff}}\) controls how low-energy observables scale under coarse graining. As \(d_{\mathrm{eff}}\) is lowered, long-wavelength collective degrees of freedom acquire greater relative weight, allowing increasingly large regions to contribute to the long range multipartite entanglement.  This scale-dependent mechanism is distinct from topologically ordered systems, where long-range entanglement is encoded in scale-independent global constraints of the ground state.

    The paper is organized as follows. In Section \ref{Section2}, we we prove that any nonvanishing multipartite entanglement quantity whose holographic prescription is constructed from a finite network of RT surfaces has the same leading large-distance power-law exponent fixed by the IR geometry. In Section \ref{Section3}, we apply this result to isotropic HV geometries, demonstrating the emergence of sub-volume-law and volume-law long-range entanglement. In Section \ref{Section4}, we investigate anisotropic HV geometries, with AdS\(_3 \times \mathbb{R}^2\) as a concrete example, and compute various multipartite entanglement quantities. Finally, Section \ref{Section5} contains our discussion and outlook.

%%%%%%%%%%%%%%%%%%%%%%%%%
\section{Universal IR Scaling for entanglement quantities}
    \label{Section2}
    In this section, we study the large-distance structure of multipartite entanglement in gapless holographic many-body states. This problem is conceptually distinct from the familiar long-range entanglement of gapped topological phases \cite{Wen2013,Levin2006,Kitaev2006}. In ordinary gapped phases, correlations are short-ranged and suppressed at long scales, so that the long-distance entanglement structure is trivial. For gapped topologically ordered states, by contrast, there can be nontrivial long-range entanglement, but it is encoded in robust, scale-independent topological data, such as the shape-independent constant term in the topological entanglement entropy. Gapless systems are different in a more fundamental way: because there is no intrinsic correlation length, the entanglement structure does not reduce at large distances to either a trivial constant or a purely topological invariant. Instead, it can retain a genuine scale dependence controlled by the IR fixed point, appearing as power-law decay or logarithmic violation. The relevant question is therefore how the quantum entanglement is organized across scales and how this scaling is determined by the IR geometry in the holographic setup. Of particular interest is the possibility that gapless systems may support nonvanishing, and in some cases scale-growing, long-distance quantum entanglement: a form of long-range entanglement that is reminiscent of gapped topological phases in its nonlocal character, but differs from them in origin, robustness, shape dependence, and physical interpretation.

    By long-distance entanglement, we do not mean the mutual information between two fixed-size subregions separated by a large distance. Such a bipartite diagnostic often decays rapidly, and in holographic systems it can vanish identically once the corresponding entanglement wedge disconnects. The relevant long-distance structure could instead be intrinsically multipartite. Two distant regions need not share a connected wedge by themselves; nevertheless, they may both participate in a larger connected entanglement wedge when the intermediate regions and the full multipartite configuration are included. We therefore focus on holographic multipartite entanglement structures for large boundary subregions, especially on scaling configurations in which the sizes and separations of all subregions are increased together so that the overall shape data are kept fixed. Such configurations provide a controlled probe of how multipartite entanglement persists, reorganizes, or decays at long distances. In this work, the multipartite entanglement measure/quantity will be understood in a broad sense, including $n$-partite information, conditional mutual information, the Markov gap, genuine multipartite entropy, and, more generally, arbitrary quantities constructed as finite linear combinations of holographic minimal-surface networks.   

    The large distance scaling of these multipartite entanglement structures is physically significant because it detects directly the IR structure of the quantum state. While the leading UV contribution to entanglement entropy is governed by the local area law and is therefore insensitive to the long-distance phase structure, the large $l$ finite behavior will be shown to be completely controlled by the deep IR geometry. The multipartite entanglement at long distance scales shares the same scaling exponents that control thermodynamics and transport, with all of them being different projections of the same underlying IR scaling structure. This makes the large-distance scaling of quantum entanglement  a possible nonlocal, order-parameter-like characterization of distinct gapless phases \cite{Baggioli2023,Ji_2025,Ju2024,Ju2026,chen2026,chen2026detectingtopologicaltransitionsanisotropy}.

    In our previous work on holographic topological semimetals, we numerically observed that several multipartite entanglement quantities exhibit power-law decay at large distances, with exponents controlled by the corresponding IR geometries. Similar behavior also appears in pure AdS, where the power law can be obtained analytically \cite{chen2026,chen2026detectingtopologicaltransitionsanisotropy}. These observations suggest a more general principle: for holographic quantities whose large-scale dependence is built from finite combinations of minimal surfaces, the leading long-distance scaling is fixed by the same IR geometric data, up to possible cancellations of leading terms. The purpose of this section is to make this statement systematic. We first review some representative multipartite entanglement quantities in hologaphy, and then show how the IR scaling in the IR geometry controls the large distance behavior of a broad class of multipartite entanglement quantities constructed from minimal-surface networks.

    \subsection{Review of holographic multipartite entanglement quantities}
        In the semiclassical large-\(N\) limit of a holographic system, the entanglement entropy of a boundary region \(A\) is determined by the Ryu--Takayanagi surface \(\gamma_A\) \cite{Ryu2006},
        \begin{equation}
            S(A)=\frac{\operatorname{Area}(\gamma_A)}{4G_N}.
        \end{equation}
        There is no unique quantity that completely characterizes multipartite entanglement. Different measures and signals probe different aspects of the underlying entanglement structure. We therefore briefly review the representative quantities relevant to the minimal-surface networks considered below.

        A basic entropy-based quantity is the \(n\)-partite information,
        \begin{equation}
            I_n(A_1:\cdots:A_n)=\sum_{\emptyset\neq J\subseteq\{1,\ldots,n\}}(-1)^{|J|+1}S(A_J),\qquad A_J=\bigcup_{j\in J}A_j .
        \end{equation}
        For a generic state, \(I_n\) is a multipartite correlation diagnostic rather than a faithful measure of genuine \(n\)-partite entanglement. Its interpretation becomes sharper in a holographic exclusive global multipartite entanglement configuration (HEGMEC). In this special type of configuration, the relevant fewer-party entanglement wedges are all disconnected, whereas the complete multiparty configuration retains the required global connectivity. The fewer-party channels are therefore excluded geometrically, and the signed quantity \(\mathcal{I}_n=(-1)^n I_n\) isolates the exclusive global multipartite contribution measured by \(I_n\). Since each entropy entering \(I_n\) is an RT area, \(I_n\) is represented by a finite linear combination of minimal surfaces.

        The multi-entropy provides a different extension of entanglement entropy to a multipartite pure state. Holographically, the tripartite multi-entropy is represented by a minimal Steiner network \cite{Harper2024},
        \begin{equation}
            S^{(3)}(A:B:C)=\frac{1}{4G_N}\min_{\{\Gamma_A,\Gamma_B,\Gamma_C\}}\sum_{X=A,B,C}\operatorname{Area}(\Gamma_X),
        \end{equation}
        where the three surfaces divide the bulk into components homologous to \(A\), \(B\), and \(C\), respectively. At a symmetric junction they obey the usual force-balance condition. The multi-entropy contains both bipartite and irreducible tripartite contributions. The latter can be isolated by the genuine tripartite multi-entropy
        \begin{equation}
            \kappa(A:B:C)=S^{(3)}(A:B:C)-\frac{1}{2}\bigl[S(AB)+S(BC)+S(CA)\bigr].
        \end{equation}
        This subtraction removes the contribution of triangle states, which are composed entirely of bipartite entanglement, while allowing \(\kappa\) to remain nonzero for genuinely tripartite states.

        For a mixed state on \(A\cup B\), the entanglement wedge cross section (EWCS) is the minimum-area surface that divides a connected entanglement wedge into parts homologous to \(A\) and \(B\) \cite{Dutta2021,Umemoto2018},
        \begin{equation}
            E_W(A:B)=\frac{1}{4G_N}\min_{\gamma_{A:B}}\operatorname{Area}(\gamma_{A:B}).
        \end{equation}
        In the semiclassical holographic limit, the standard relations are \(E_P(A:B)=E_W(A:B)\) and \(S_R(A:B)=2E_W(A:B)\), where \(E_P\) and \(S_R\) denote the entanglement of purification and reflected entropy, respectively. The Markov gap is then \cite{Hayden2021,Zou2021}
        \begin{equation}
            h(A:B)=S_R(A:B)-I(A:B)=2E_W(A:B)-I(A:B).
        \end{equation}
        It measures the part of the reflected entropy not accounted for by mutual information. A nonzero Markov gap excludes a sum-of-triangle-states structure and therefore signals irreducible tripartite entanglement. However, it is not a complete genuine multipartite entanglement quantity, since it can vanish for generalized GHZ states. A complementary quantity constructed from the upper bound of reflected entropy is the bipartite latent entropy,
        \begin{equation}
            \ell_{AB}=2\min\{S(A),S(B)\}-S_R(A:B).
        \end{equation}
        For an \(n\)-partite pure state, its multipartite extension is the geometric mean
        \begin{equation}
            \ell_{A_1\cdots A_n}=\left(\prod_{i<j}\ell_{A_iA_j}\right)^{\frac{2}{n(n-1)}} .
        \end{equation}
        The L-entropy vanishes on states separable across a bipartition and is sensitive to both GHZ- and W-type entanglement, thereby complementing the Markov gap.

        The EWCS also admits a multipartite generalization. For a connected entanglement wedge of \(A_1\cup\cdots\cup A_n\), the multi-EWCS is defined by \cite{Bao2019,Bao2019Another}
        \begin{equation}
            E_W(A_1:\cdots:A_n)=\frac{1}{4G_N}\min_{\{\Gamma_i\}}\operatorname{Area}\left(\bigcup_{i=1}^{n}\Gamma_i\right),
        \end{equation}
        where the cross sections partition the wedge into components homologous to the corresponding boundary regions. It is the holographic counterpart of multipartite entanglement of purification. For three finite regions \(A,B,C\), the complement \(D=\overline{A\cup B\cup C}\) acts as an implicit purifier, so that \(E_W(A:B:C)\) probes a four-partite structure. Two associated signals are
        \begin{align}
            \Delta_W(A:B:C)&=E_W(A:B:C)-E_W(A:BC)-E_W(B:AC)-E_W(C:AB),\\
            g_W(A:B:C)&=E_W(A:B:C)-I_{\mathrm{tot}}(A:B:C),
        \end{align}
        where \(I_{\mathrm{tot}}(A:B:C)=S(A)+S(B)+S(C)-S(ABC)\). The first subtracts contributions already captured by bipartite partitions, whereas the second is a multipartite analogue of the Markov gap.

        The quantities reviewed above include linear combinations of RT areas, Steiner-type networks, and cross sections of connected entanglement wedges. Despite their different information-theoretic interpretations, their holographic representations are all finite networks of bulk minimal surfaces. This common geometric structure is the basis for the unified IR scaling analysis in the following subsection.

   %\textcolor{cyan}{We start from prove that for a strip region of width $l_i$ aligned along the $x^i$ direction on the boundary, the turning point $r_*$ of the corresponding RT surface satisfies $l_i \propto r_*^{\,1-\alpha_i+\alpha_r}$, which establishes the power-law relation that will be crucial for the subsequent proof. We then demonstrate that for any entanglement measure $\mathcal{Q}$ constructed from combinations of RT surfaces for such strip regions, its scaling under a boundary scale transformation $x^i \to \lambda x^i$ is}
    %\begin{equation}
    %    \mathcal{Q} \propto \lambda^{\,1+\frac{\alpha_V}{1-\alpha_i+\alpha_r}},
    %\end{equation}
    %where $\alpha_r$, $\alpha_i$, and $\alpha_V$ are constants determined by the leading power-law exponents of the IR metric. Throughout this section we assume $1-\alpha_i+\alpha_r \neq 0$.

\subsection{Long range multipartite entanglement from the IR geometry}
    {We focus on holographic zero-temperature quantum systems without the presence of horizons, and the bulk metric is taken to be asymptotically AdS. We consider a static spacetime and the boundary field theory is equipped with translational symmetry; accordingly, on a suitable time slice, all metric components depend only on the radial coordinate $r$. For simplicity, we choose suitable bulk coordinates $\{x^1,x^2,\dots,x^d,r\}$ in $d$-spatial dimensions, in which the metric for the bulk time slice is diagonalized as\footnote{In fact, the presence of off-diagonal terms in the metric does not affect the final conclusions.}}
    %On a chosen Cauchy slice, we can always choose a suitable set of coordinates $\{x^1,x^2,\dots,x^n,r\}$ such that the metric is diagonalized as
    \begin{equation}
        \label{Metric}
        \sum_{i=1}^d g_{ii}\,(dx^i)^2 + g_{rr}\,(dr)^2.
    \end{equation}
    {In the case of a pure AdS bulk, all $g_{ii}$ are equal to $r^2$ and $g_{rr}=1/r^2$. %In the general case, however, they are determined by the IR geometry. 
    For most systems, due to anisotropy, $g_{ii}$ could be different functions in different directions $i$. As $r\to\infty$ approaching the AdS boundary, all $g_{ii}$ tend to $r^2$ and $g_{rr}$ tends to $1/r^2$.} Denote $f_i = g_{ii}$, $f_V = \prod_{i=1}^d g_{ii}$, and $f_r = g_{rr}$. 
    
    The low-energy, large-scale behavior of the system is governed by the IR geometry. Since holographic systems typically describe gapless systems, which possess no characteristic scale, they generally exhibit scaling symmetry at the IR fixed point. Therefore, we consider systems that possess scaling symmetry at the IR fixed point. In the deep IR regime, i.e. in the limit $r\to 0$, the metric is approximated by its leading-order power-law behavior
    \begin{equation}
        \label{PowerLaw}
        f_i \propto r^{2\alpha_i},\qquad f_V \propto r^{2\alpha_V},\qquad f_r \propto r^{2\alpha_r},
    \end{equation}
    where the exponents $\alpha_r$, $\alpha_i$, and $\alpha_V$ completely determine the IR scaling behavior. {Here the exponents $\alpha_r$, $\alpha_i$, $\alpha_V$ are invariant under the scaling transformation $r\to \lambda r$, but they are not necessarily invariant under more general reparameterizations. Nevertheless, we can ensure that the final result only depends on a gauge invariant combination of these exponents.}

    {To examine the distance dependence of entanglement measures, one may employ strip configurations, since strips effectively isolate the distance dependence of quantum entanglement by eliminating irrelevant shape dependence, thereby allowing us to focus purely on the long-range behavior along the strip width direction. Owing to the translation symmetry of the boundary field theory, the RT surface associated with a strip region on the boundary is always completely integrable. Without loss of generality, for a strip of width $l_i$ along the $x^i$ direction, if its RT surface is homeomorphic to the strip region on the boundary, which means the RT surface is a connected minimal surface, then its cross profile in the $x^i$ direction takes the shape of an arch.

    The point deepest in the IR region is denoted as the turning point $r_*$, where $\frac{dr}{dx^i}(r_*)=0$. This point reflects, to some extent, how deeply the RT surface probes the IR geometry, and correspondingly, how much the entanglement quantity of the boundary region captures the long-range properties. Moreover, $r_*$ will play a crucial role in the subsequent calculations. Specifically, for this type of RT surface, the turning point depth $r_*$ and the strip width $l_i$ uniquely determine each other, and the relation is}
    \begin{equation}
        \label{StripWidthIntegral}
        l_i(r_*) = 2\int_{r_*}^{\infty} \sqrt{\frac{f_r(r)\,C_i^2}{f_i(r)\bigl(f_V(r)-C_i^2\bigr)}}\,dr.
    \end{equation}
    %$r_*$ is the turning point where $r'(r_*)=0$ and 
    where $C_i = \sqrt{f_V(r_*)}$ is the derivative for entanglement $S_i$ corresponding to the strip, i.e. $C_i=dS_i/dl_i$. If the functions $f_i$, $f_r$, and $f_V$ in \eqref{StripWidthIntegral} obey the power-law behavior specified in \eqref{PowerLaw} throughout the entire region $r>0$, the integrand can be integrated analytically, yielding the exact power law
    \begin{equation}
        \label{StripWidthPowerLaw}
        l_i = \frac{\sqrt{\pi}}{\alpha_V}\frac{\Gamma\left(\frac{1}{2}-\frac{1-\alpha_i+\alpha_r}{2 \alpha_V}\right)}{\Gamma\left(1-\frac{1-\alpha_i+\alpha_r}{2 \alpha_V}\right)}\; r_*^{1+\alpha_r-\alpha_i}.
    \end{equation}
    Note that we focus on bulk metric that is asymptotically AdS. Consequently, a global power-law metric of \eqref{PowerLaw} cannot hold throughout the entire bulk. Nonetheless, for the long-range physics that we are interested in, the power-law parametrization \eqref{StripWidthPowerLaw} remains a useful approximation and will be employed in the subsequent proof.

    For a general metric we consider in \eqref{Metric}, the IR region ($r\to 0$) is approximately described by power laws in \eqref{PowerLaw}. When the width $l_i$ of the strip region in the boundary field theory is large, the turning point $r_*$ of the corresponding RT surface tends to zero. To show that when $r_*\to 0$ the integral \eqref{StripWidthIntegral} is dominated by the deep IR part (so that the power law \eqref{StripWidthPowerLaw} holds asymptotically), we use the dominated convergence theorem.

    Denote the integrand
    \begin{equation}
        \nonumber
        f_{r_*}(r) = 2\sqrt{\frac{f_r(r)\,C_i^2}{f_i(r)\bigl(f_V(r)-C_i^2\bigr)}},\qquad r\ge r_*.
    \end{equation}
    Given a sufficiently small $\delta>0$ so that the region $r<\delta$ is sufficiently deep in the IR, where the metric is dominated by the leading-order term, we show that the integral receives its dominant contribution from the interval $[r_*,\delta]$, while the tail $\int_{\delta}^{\infty} f_{r_*}(r)\,dr$ vanishes as $r_*\to 0$, which corresponds to the boundary width $l_i$ tending to infinity. The proof is as follows.
    
    We construct an $r_*$-independent dominant function $g(r)$ such that $0\le f_{r_*}(r)\le g(r)$ for all sufficiently small $r_*$, with $g(r)$ integrable on $[\delta,\infty)$. Note that for fixed $r\ge\delta$, the function $r_*\mapsto f_{r_*}(r)$ is monotonic (typically decreasing as $r_*$ decreases, because $C_i=\sqrt{f_V(r_*)}$ becomes smaller, making the denominator $f_V(r)-C_i^2$ larger). Consequently, for any $r_*<\delta$ we have $f_{r_*}(r)\le f_{\delta}(r)$. Hence we may simply choose
    \begin{equation}
        \nonumber
        g(r) := f_{\delta}(r) = 2\sqrt{\frac{f_r(r)\,C_i(\delta)^2}{f_i(r)\bigl(f_V(r)-C_i(\delta)^2\bigr)}},\qquad r\ge\delta,
    \end{equation}
    which is integrable on $[\delta,\infty)$.

    As $r_*\to 0$, we have $C_i=\sqrt{f_V(r_*)}\to 0$, hence for every fixed $r\ge\delta$, we have $\lim_{r_*\to 0}f_{r_*}(r) = 0$. By the dominated convergence theorem,
    \begin{equation}
        \nonumber
        \lim_{r_*\to 0}\int_{\delta}^{\infty} f_{r_*}(r)\,dr = 0.
    \end{equation}

    Take $\delta$ sufficiently small so that the interval $[r_*,\delta]$ lies deep inside the IR region where the metric is essentially power-law. Then decompose
    \begin{equation}
        \nonumber
        l_i(r_*) = \int_{r_*}^{\delta} f_{r_*}(r)\,dr \;+\; \int_{\delta}^{\infty} f_{r_*}(r)\,dr.
    \end{equation}
    As $r_*\to 0$, the tail integral vanishes. The main contribution $\int_{r_*}^{\delta} f_{r_*}(r)\,dr$, under the IR power-law approximation, yields exactly the power-law form \eqref{StripWidthPowerLaw} (the constant factor may receive small corrections from the finite $\delta$, but the exponent remains unchanged). Therefore, for sufficiently small $r_*$,
    \begin{equation}
        \nonumber
        l_i \propto r_*^{1-\alpha_i+\alpha_r},
    \end{equation}
    which is the desired power-law relation dictated by the IR scaling exponents. {This conclusion has been numerically confirmed in previous work \cite{chen2026}.}

    The case $1-\alpha_i+\alpha_r>0$ is unphysical, because in that case, if the RT surface corresponding to the strip region on the boundary remains homeomorphic to the boundary strip region, then as the strip width $l_i$ increases, the turning point depth $r_*$ would instead increase. This would imply that two different RT surfaces intersect in two separate curves, which contradicts the general theory of ordinary differential equations. Therefore, we exclude this unphysical scenario. When $1-\alpha_i+\alpha_r=0$, the relation $l_i \propto r_*^{1-\alpha_i+\alpha_r}$ would imply $l_i \propto r_*^0$ at large $l_i$, suggesting that $l_i$ becomes a constant independent of $r_*$; this is clearly unphysical for gapless systems. In fact, it is not difficult to see that when $1-\alpha_i+\alpha_r=0$, the integral in \eqref{StripWidthIntegral} has an upper bound in $r_*$, so there is no guarantee that for every $l_i$ there exists a minimal surface homeomorphic to the strip. This indicates that a phase transition occurs as $l_i$ increases: the shape of the RT surface changes from being homeomorphic to the strip on the boundary to two walls extending from the strip edges to the horizon. Such a transition typically indicates that the system becomes gapped in the $x^i$ direction, with the corresponding low-energy excitations being suppressed. This corresponds to the entanglement entropy between the strip and its complement transitioning to a pure area‑law for this gapped system. 
%%%%%%%%%%%%%%%%%%%%%%%%%
%\section{the proof of $\mathcal{Q} \propto l_i^{\,1+\frac{\alpha_V}{1-\alpha_i+\alpha_r}}$}
    \begin{figure}[htbp]
        \centering

        \begin{minipage}[b]{0.48\textwidth}
            \centering
            \includegraphics[width=\linewidth]{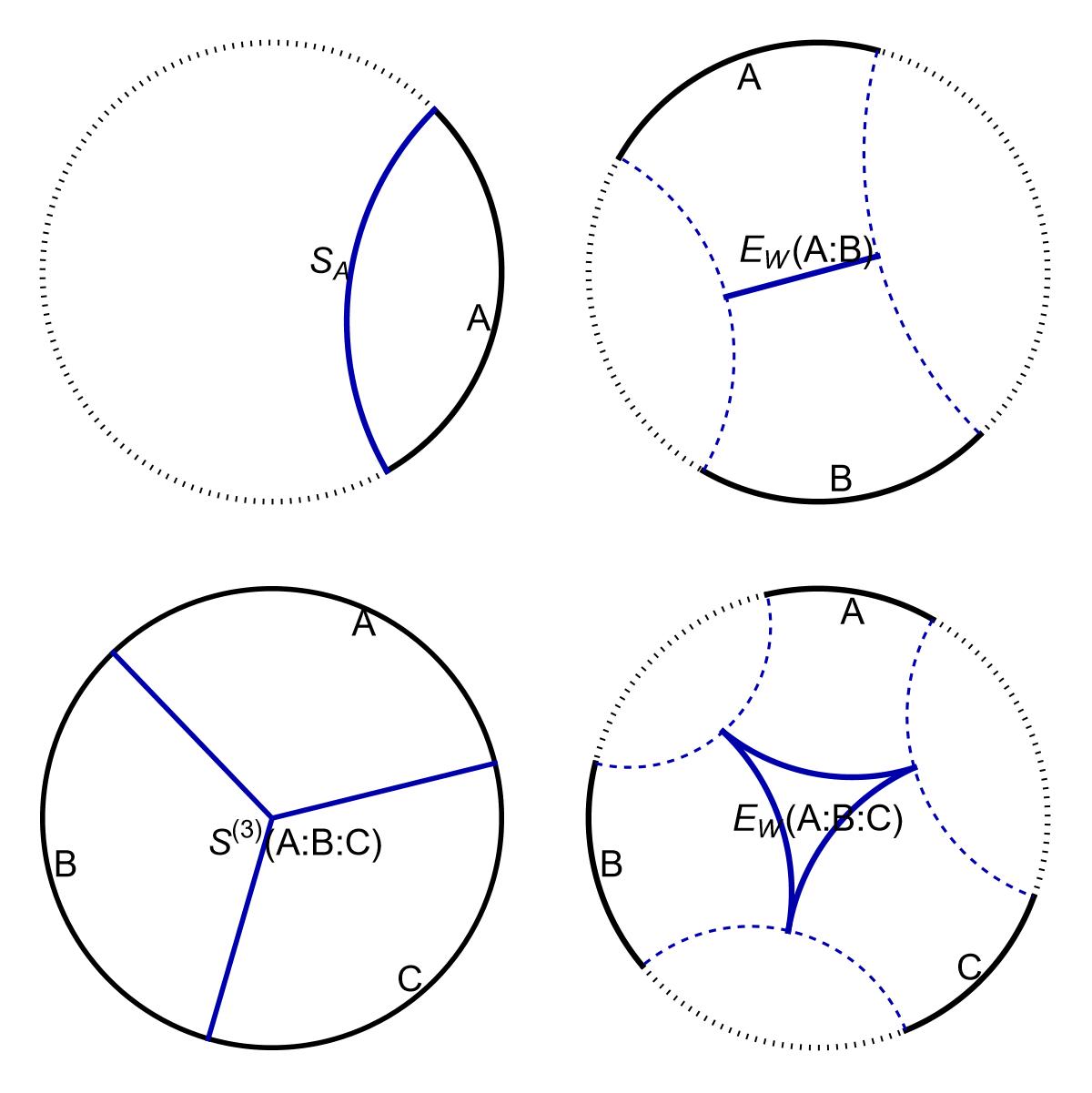}
        \end{minipage}
        \hfill
        \begin{minipage}[b]{0.48\textwidth}
            \centering
            \includegraphics[width=\linewidth]{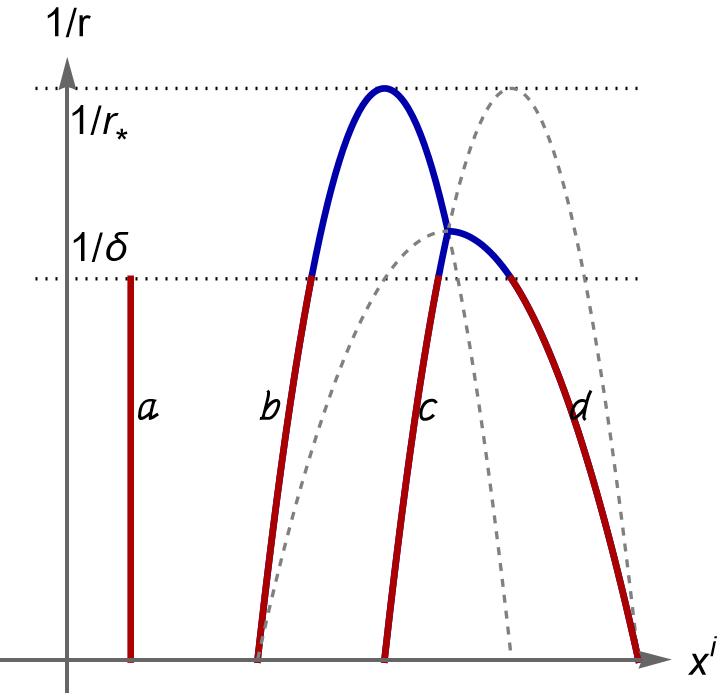}
        \end{minipage}
        \caption{Left panel: On the Poincaré disk, we display the minimal surface networks corresponding to several typical entanglement measures, namely the entanglement entropy $S_A$ for region $A$, the entanglement wedge cross-section $E_W(A:B)$ for regions $A$ and $B$, the tripartite multi-entropy $S^{(3)}(A:B:C)$ and the tripartite multi-EWCS $E_W(A:B:C)$ for subregions $A,B,C$. $\qquad$
        Right panel: On the Poincaré upper half-plane, we show a typical minimal surface network, where each minimal surface is part of some RT surface, completed by gray dashed lines in the figure. By choosing a sufficiently small $\delta>0$, one can ensure that the metric above the horizontal line at $1/\delta$ is approximately power-law. It is argued that all parts of the minimal surfaces below the horizontal line at $1/\delta$, marked in red, have equal areas. Note that the ``straight leg'' minimal surface labeled $a$ differs from $b,c,d$ in that the $x^i$-coordinates of surface $a$ are constant, implying that it carries no information about the $l_i$ scaling and thus does not contribute any non-trivial term to the scaling behavior of the entanglement measures.}
        \label{FigureNetWork}
    \end{figure}

    Next we focus on the case of \(1-\alpha_i+\alpha_r<0\) with $r_* \propto l_i^{\frac{1}{1-\alpha_i+\alpha_r}}$. From this, one can study the large‑scale behavior of various strip‑based entanglement quantities. We now investigate the scaling behavior of generic multipartite entanglement quantities constructed from combinations of various minimal surfaces. Such minimal surface networks can be quite complicated, and the minimal surfaces within them do not necessarily extend all the way to the AdS boundary and remain homeomorphic to the strip region on the boundary as RT surfaces do. Instead, they can serve as minimal cross-sections of the entanglement wedge, like the EWCS, or form Steiner trees as required in the context of multi-entropy. It suffices that they satisfy the minimal surface equation everywhere except at the network nodes. It is not difficult to see that, in this case, the minimal surfaces in the network are either, as shown in the left panel of Fig.\,\ref{FigureNetWork}, part of an RT surface corresponding to a strip region on the boundary, taking the shape of an arch, or are constant in every $x^i$ direction, taking the shape of a vertical wall. 
    
    {To probe the long-range scaling behavior of multipartite entanglement, we consider configurations in which all subregion lengths are taken to be large. In order to compare configurations at different large scales while keeping the relative proportions among all subregions fixed, we require that they expand at the same rate. The most convenient way to achieve this is to introduce a global scaling transformation that acts on the entire configuration simultaneously, so that all scale dependence becomes the dependence on a single parameter $\lambda$. For configurations composed of strips oriented along the $x^i$ direction, we define this scaling transformation by a factor $\lambda$ as the mapping}
    \begin{equation}
        x^i \mapsto \lambda x^i.
    \end{equation}

    It is easy to see that no matter how complicated the configuration on which the entanglement quantity depends, under the scaling transformation $\lambda\to\infty$, its intricate network structure, turning points, and nodes all penetrate deep into the IR region. We shall next show that at large scales, the part outside the IR region contributes only simple constant and divergent behaviors that are independent of both the configuration and the boundary scale. 
    
    Choose \(\delta\) sufficiently small so that it lies deep in the IR region. We first show that as \(r_* \to 0\), the following integral tends to \(0\):
    \begin{equation}
        \label{StraightLeg}
        \int_{\delta}^\infty\left(\prod_{j=1}^d g_{jj}\sqrt{\frac{g_{rr}}{g_{ii}(\prod_{j=1}^d g_{jj}-C_i^2)}}-\sqrt{g_{rr}\prod_{j\ne i}g_{jj}}\right)dr,
    \end{equation}
    {where the first integral term in the integrand corresponds to the area of minimal surfaces of the type $b,c,d$ in the network configuration as shown in the right panel of Fig.\,\ref{FigureNetWork}, while the second integral term corresponds to the area of the ``straight leg'' type minimal surface labeled $a$ in Fig.\,\ref{FigureNetWork}. We only need to show that the difference between the two tends to zero in the large-scale limit, which implies that in the region $r>\delta$, the segments of the minimal surfaces all approach the straight-leg type, carrying no scaling information and contributing only scale-independent constant and divergent terms. This will create favorable conditions for our subsequent treatment.} The integrand for \eqref{StraightLeg} is positive, integrable, and converges pointwise to \(0\). To apply the dominated convergence theorem, we need a fixed integrable dominating function. Denote the integrand by \(h_{r_*}(r)\). Since for fixed \(r\), \(h_{r_*}(r)\) is monotonically decreasing in \(r_*\), we may choose a sufficiently small \(r_*^{(0)}\) (e.g., \(r_*^{(0)}<\delta\)) such that for all \(r_*<r_*^{(0)}\),
    \begin{equation}
        \nonumber
        h_{r_*}(r) \le h_{r_*^{(0)}}(r).
    \end{equation}
    The function \(h_{r_*^{(0)}}(r)\) is fixed and integrable. Hence, the dominated convergence theorem ensures that the \eqref{StraightLeg} tends to \(0\).

    Therefore, as \(r_*\to 0\), the contribution from the region \(r>\delta\) becomes negligible. In this region, the minimal surface reduces to a “straight leg” structure independent of \(l_i\), whose area contributes only a constant or divergent term that does not affect the leading power-law behavior in $l_i$. Thus, in analyzing the large-scale scaling behavior, we only need to focus on the deep IR region \([r_*, \delta]\). Consequently, we may effectively replace the metric for \(r>\delta\) by the same power-law form as in the deep IR
    \begin{equation}
        g_{ii}\sim r^{2\alpha_i},\qquad g_{rr}\sim r^{2\alpha_r}.
    \end{equation}
    As argued above, this replacement affects only the constant and divergent contributions, not the part we are interested in.

    Next, we perform the rescaling
     \begin{equation} 
        \widetilde{x^i}=\lambda^{-1} x^i,\qquad \widetilde{x^j}=x^j,\qquad \widetilde{r}=\lambda^{-\frac{1}{1-\alpha_i+\alpha_r}} r.
     \end{equation}
%    Under this rescaling, the metric transforms as
%    \begin{equation}
%        \nonumber
%        \sum_{j=1}^n\widetilde{g}_{jj}(d\widetilde{x^j})^2+\widetilde{g}_{rr}(d\widetilde{r})^2=\lambda^2 g_{ii}(d\widetilde{x^i})^2+\sum_{j\ne i} g_{jj}(d\widetilde{x^j})^2+\lambda^{\frac{2}{1-\alpha_i+\alpha_r}}g_{rr}(d\widetilde{r})^2,
%    \end{equation}
%    or explicitly,
%    \begin{equation}
%        \nonumber
%        {\lambda^2}{r^{2\alpha_i}}(d\widetilde{x^i})^2+\sum_{j\ne i} r^{2\alpha_j}(d\widetilde{x^j})^2+\lambda^{\frac{2}{1-\alpha_i+\alpha_r}}r^{2\alpha_r}(d\widetilde{r})^2.
%    \end{equation}
    
    In the coordinates \((\widetilde{x^i},\widetilde{r})\), the shape of the minimal surface network remains invariant. {More specifically, as the boundary scale $\lambda$ increases, the coordinate components of points on the minimal surface network in the local coordinates $(\widetilde{x^i},\widetilde{r})$ remain invariant.} It suffices to check that the following expression is independent of \(\lambda\):
    \begin{equation}
        \widetilde{x^i}_2-\widetilde{x^i}_1=\int_{\widetilde{r}_1}^{\widetilde{r}_2}\sqrt{\frac{{g}_{rr}(\widetilde{r})\,\widetilde{C_i}^2}{{g}_{ii}(\widetilde{r})\bigl(\prod_{j=1}^d{{g}_{jj}(\widetilde{r})}-\widetilde{C_i}^2\bigr)}}\,d\widetilde{r},
    \end{equation}
    where \(\widetilde{C_i}=\sqrt{\prod_{j=1}^n {g}_{jj}(\widetilde{r}_*)}\) and $\widetilde{r}_*=\lambda^{-\frac{1}{1-\alpha_i+\alpha_r}} r_*$. This can be verified directly. The verification procedure is relatively lengthy and contains no particular subtleties; hence, it is omitted here.

    We now examine how the area of any shape-invariant minimal surface changes under the above scaling transformation. {As shown in the right panel of Fig.\,\ref{FigureNetWork}, there are two different types of minimal surfaces: one type, such as $b,c,d$, is part of an RT surface, whose area we denote as $S_1$; the other type, as illustrated by $a$, is of the ``straight leg'' type, whose area we denote as $S_2$. We need to verify the behavior of $S_1$ and $S_2$ under scaling transformations separately. Since we have already established that the shape of the minimal surface network is invariant under the coordinates $(\widetilde{x^i},\widetilde{r})$, we can directly examine the integrals for $S_1$ and $S_2$ in these coordinates. For a minimal surface in a minimal surface network, if the radial coordinates of its two endpoints are $r_1$ and $r_2$ with $r_1<r_2$ in the absence of a scaling transformation (i.e., at $\lambda=1$), then for both types of minimal surfaces, their areas are directly determined by the coordinates on the interval $[r_1,r_2]$.}
    \begin{equation}
        \begin{aligned}
            S_1&=\int_{r_1}^{r_2}\prod_{j=1}^d g_{jj}\sqrt{\frac{g_{rr}}{g_{ii}(\prod_{j=1}^d g_{jj}-C_i^2)}}\,dr,\\
            S_2&=\int_{r_1}^{r_2}\sqrt{g_{rr}\prod_{j\ne i}g_{jj}}\,dr.
        \end{aligned}
    \end{equation}
    {After performing a scaling transformation with $\lambda>1$, the span of the same minimal surface segment along the $x^i$ direction changes from $[x^i_1, x^i_2]$ to $[\lambda x^i_1, \lambda x^i_2]$. From our previous conclusion, under the scaling transformation, the shape of the minimal surface remains invariant in the coordinates $(\widetilde{x}^i, \widetilde{r})$, and the radial coordinates of the two endpoints change from $r_1, r_2$ to $\lambda^{\frac{1}{1-\alpha_i+\alpha_r}} r_1, \lambda^{\frac{1}{1-\alpha_i+\alpha_r}} r_2$. Accordingly, the areas of the corresponding minimal surfaces are given by}
    \begin{equation}
        \begin{aligned}
            \widetilde{S}_1 &= \int_{\widetilde{r}_1}^{\widetilde{r}_2} \prod_{j=1}^d g_{jj}(\widetilde{r}) \sqrt{ \frac{g_{rr}(\widetilde{r})}{ g_{ii}(\widetilde{r}) \bigl( \prod_{j=1}^d g_{jj}(\widetilde{r}) - \widetilde{C}_i^2 \bigr) } } \, d\widetilde{r}, \\
            \widetilde{S}_2 &= \int_{\widetilde{r}_1}^{\widetilde{r}_2} \sqrt{ g_{rr}(\widetilde{r}) \prod_{j\ne i} g_{jj}(\widetilde{r}) } \, d\widetilde{r}.
        \end{aligned}
    \end{equation}
    Finally, it is straightforward to show that for both types of minimal surfaces, their scaling behavior is the same: 
    \begin{equation}
        \widetilde{S}=\lambda^{1+\frac{\alpha_V}{1-\alpha_i+\alpha_r}}\,S,
    \end{equation}
    where \(\alpha_V=\sum_j \alpha_j\). We thus arrive at the final conclusion: any entanglement quantity $\mathcal{Q}$  constructed directly from combinations of various types of minimal surfaces for strip configurations without {\it leading cancellation} exhibits a universal power-law scaling at large scales
    \begin{equation}
        \label{Result}
        \boxed{\mathcal{Q} \propto \lambda^{\,1+\frac{\alpha_V}{1-\alpha_i+\alpha_r}} }.
    \end{equation}
    This conclusion encompasses all entanglement quantities defined by holographic minimal surfaces, such as entanglement entropy, mutual information, tripartite mutual information, entanglement wedge cross sections (EWCS), and multi-EWCS.

    As we have mentioned earlier, although the exponents $\alpha_i$, $\alpha_r$, and $\alpha_V$ are not necessarily invariant under reparameterizations, the combination in \eqref{Result} is gauge invariant. This guarantees that our conclusions are gauge-invariant and hence physically sensible. Note that \eqref{Result} is valid for an arbitrary number $n$ of parties, including tripartite, four-partite, five-partite, and higher-partite entanglement quantities. Thus we have actually established that in holographic quantum many-body systems, as long as a certain type of multipartite entanglement quantity is non-zero, it will obey the same scaling behavior as other non-zero multipartite entanglement quantities. This implies that once a particular type of entanglement exists, its strength is comparable to that of other types, reflecting the complexity and diversity of the entanglement structure in holographic systems.

    Moreover, the large-scale behavior of these entanglement quantities provides a significant step forward in reconstructing the bulk geometry of holographic systems. While the UV part of the geometry has been relatively accessible in previous studies, the IR geometry-especially its leading power-law scaling-can now be directly read off from the scaling behavior of the entanglement quantities. This leading exponent is gauge-invariant and thus carries unambiguous physical information. The leading coefficient of the IR geometry, however, is sensitive to the crossover region that interpolates between the IR and UV regimes; this transition region dictates how the two asymptotic geometries are connected. Therefore, by fixing the gauge, one can also determine this leading coefficient from the entanglement data, thereby completing the reconstruction of the IR geometry up to the leading order. The observation that the large-scale scaling behavior of boundary entanglement quantities allows for a direct reconstruction of the IR geometry lends further support to our view that entanglement is the more fundamental quantity, and that the bulk geometry is emergent from the entanglement structure of the boundary field theory.
%%%%%%%%%%%%%%%%%%%%%%%%%
\section{Long-Range Entanglement in holographic gapless systems}
    \label{Section3}

    The purpose of this section is to apply the general scaling result of Section \ref{Section2} to a concrete class of IR geometries and to identify the conditions under which long-distance multipartite entanglement can be enhanced. Since the leading scaling of entanglement quantities as geometrized via RT networks is governed by the IR geometry at large distances, the key question is whether a realistic holographic system can admit an IR geometry that genuinely supports a large amount of long-range entanglement. HV geometries provide a particularly useful setting. HV geometries constitute a broad family of scale-covariant geometries characterized by a variety of tunable scaling exponents, and therefore support many nontrivial scaling behaviors of thermodynamic, transport, and entanglement observables. In this work, we take the HV geometry to describe only the deep IR region, while the full bulk geometry remains asymptotically AdS$_5$ in the UV.
   
    Our choice of HV geometries is motivated by our previous study of long-range entanglement behavior as a nonlocal probe of quantum phase transitions in holographic topological systems, including holographic Weyl semimetals \cite{Landsteiner2016}, holographic topological nodal line semimetals \cite{Liu2021}, holographic Weyl-\(\mathbb{Z}_2\) semimetals \cite{Ji2021}, and holographic Weyl-nodal line coexisting semimetals \cite{Chu2024}. In that study, we found that the IR geometry completely determines the long-distance scaling behavior of all the multipartite entanglement quantities considered. In these examples, the leading IR geometries can be naturally organized as special limits of general anisotropic HV-type geometries with specific choices, or limiting values, of the scaling exponents \((z_t,z_i)\) and the hyperscaling violation exponent \(\theta\). HV geometries therefore provide a unified framework for organizing the long-distance entanglement behavior observed in these systems and for systematically exploring stronger long-range multipartite entanglement structures. We next demonstrate that, within specific parameter regimes, a system with the HV geometry serving as the IR geometry can significantly enhance long-range entanglement behavior.

\subsection{HV geometry as the IR geometry}

    {For a \((d+2)\)-dimensional bulk spacetime, a general anisotropic hyperscaling-violating geometry can be written as
    \begin{equation}
        \label{HyperscalingViolatingGeometry}
        ds^2=r^{-\frac{2\theta}{d}}\left(-r^{2z_t}dt^2+ \frac{dr^2}{r^2}+ \sum_{i=1}^{d}r^{2z_i}(dx^i)^2\right).
    \end{equation}
    Here \(r\) is the holographic radial coordinate, with the deep IR located at \(r\to 0\), \(t\) is the boundary time coordinate, and \(x^i\) denote the \(d\) boundary spatial coordinates. The exponent \(z_t\) characterizes the scaling of time, while \(z_i\) characterizes the generally anisotropic scaling of the \(i\)-th spatial direction. The parameter \(\theta\) is the hyperscaling-violation exponent and controls the overall Weyl scaling of the metric.

    Under the scaling transformation
    \begin{equation}
        \nonumber
        r\to s^{-1}r,
        \qquad
        x^i\to s^{z_i}x^i,
        \qquad
        t\to s^{z_t}t,
    \end{equation}
    the line element transforms covariantly as \(ds^2\to s^{\frac{2\theta}{d}}ds^2\). Thus, for \(\theta\neq 0\), the metric is scale covariant rather than strictly scale invariant. The exponents \(z_t\) and \(z_i\) determine the relative scaling between the time and spatial directions, while \(\theta\) modifies the effective scaling dimension of the IR degrees of freedom. In the general anisotropic case, the spatial volume scales with the effective exponent \(\sum_{i=1}^{d}z_i-\theta\).

    The relation to ordinary scale-invariant geometries is transparent in the limit \(\theta=0\), where the overall hyperscaling-violating Weyl factor disappears. If, in addition, \(z_t=z_i=1\), the scaling becomes relativistic and the metric reduces to pure \(AdS_{d+2}\). In the following, we start from the spatially isotropic subclass by setting all \(z_i=1\). In this case, \(z_t\) becomes the usual dynamical critical exponent \(z\), and the effective spatial dimension reduces to
    \[d_{\mathrm{eff}}=d-\theta.\]
    Consequently, variations in long-range entanglement are directly tied to changes in the effective scaling dimension, or equivalently in the number of IR degrees of freedom that remain active at long distances.}
    
    Taking the HV geometry as the IR geometry, the effective dimension $d_\text{eff}$ captures the scaling behavior of the system at low energies, namely the dimension of the degrees of freedom that remain active in the IR. More generally, in various physical systems the effective dimension \(d_{\text{eff}}\) replaces the true geometric dimension \(d\) in determining the scaling of many observables. For example, in the isotropic case, the power-law behavior of the thermal entropy density is \(s \sim T^{\frac{d_{\text{eff}}}{z_t}}\), and that of the charge density is \(\rho \sim \mu^{\frac{d_{\text{eff}}}{z_t}}\), where \(T\) is the temperature and \(\mu\) is the chemical potential. Thus, \(d_{\text{eff}}\) plays the role of an effective spatial dimension that governs the low-energy scaling physics. From a physical point of view, we always require $d_{\text{eff}}\ge 0$, i.e. $d \ge \theta$.

    Besides, we also require additional constraints on the parameters from the Null Energy Condition(NEC). Verifying the NEC for the metric \eqref{HyperscalingViolatingGeometry} yields
    \begin{equation}
        \label{NullEnergyCondition}
        \begin{aligned}
            \sum_{i=1}^d \left(z_i - \frac{\theta}{d}\right)\left(z_t - z_i - \frac{\theta}{d}\right) &\ge 0,\\
            (z_t - z_i)\left(z_t + \sum_{j=1}^d z_j - \theta\right) &\ge 0.
        \end{aligned}
    \end{equation}
    For an isotropic system, \(z_i = 1\), and these inequalities reduce to \cite{Dong2012}
    \begin{equation}
        \begin{aligned}
            (d-\theta)\bigl(d(z_t-1)-\theta\bigr) &\ge 0,\\
            (z_t-1)(d+z_t-\theta) &\ge 0.
        \end{aligned}
    \end{equation}

\subsection{Long range entanglement behavior in this geometry}
    We now investigate the scaling behavior of entanglement measures constructed from minimal surface networks when the HV geometry is taken as the IR geometry. By choosing appropriate $z_t$ and $\theta\le d$, the exponent of the radial metric component becomes $\alpha_r = -1-\frac{\theta}{d}$, and the exponents for all spatial directions are equal: $\alpha_i = 1-\frac{\theta}{d}$.

    An interesting feature of hyperscaling violating geometries is that, under this time slicing, regardless of the value of $z$, the turning point $r_*$ and the strip width $l_i$ satisfy $r_* \propto l_i^{-1}$ at large scales. Consequently, the corresponding conserved quantity $C_i = \frac{dS_i}{dl_i} \propto l_i^{-(d-\theta)}=l_i^{-d_{\text{eff}}}$. The corresponding large scaling behavior of the entanglement entropy is logarithmic when $d_{\text{eff}}=1$. When $d_{\text{eff}}>1$, the corresponding entanglement behavior exhibits a power-law decay. In other cases, $S_i \propto l_i^{1-d_{\text{eff}}}$. In particular, when $d-1 < \theta < d$, i.e., $0<d_{\text{eff}} < 1$, the entanglement entropy exhibits a sub-volume-law long-range behavior. Note that for UV divergent entanglement quantities, we consider only their $l$-dependence terms, which are IR terms without UV divergence. This is because the entanglement configuration we choose is based on strip subregions, therefore, it is guaranteed that area law UV divergent terms are independent of $l$. From the previous result~\eqref{Result}, we know that in this case any entanglement quantity constructed from combinations of RT surfaces corresponding to strips exhibits the same scaling behavior in $l$ as the IR term of the entanglement entropy. We summarize the scaling behaviors of entanglement measures and the thermal entropy for different HV parameters in Table~\ref{Table1}.

    {The scaling behaviors of entanglement quantities summarized in Table~\ref{Table1} apply to the large-scale regime, in which the corresponding minimal-surface networks extend deep into the IR geometry. At sufficiently small scales, by contrast, these networks remain in the asymptotically AdS UV region, where their leading contributions obey the usual area law. The behavior at intermediate scales is generally nonuniversal and depends on the detailed RG flow interpolating between the UV and IR fixed points. Determining this crossover behavior therefore requires solving the bulk equations of motion for a complete holographic geometry that is asymptotically AdS in the UV and approaches the HV geometry in the IR.} Explicit interpolating solutions of this type have been constructed in both bottom-up Einstein--Maxwell--dilaton models and top-down string-theory settings~\cite{Iizuka_2013,Cremonini_2016}. In this work, however, we focus only on the universal long-distance behavior governed by the IR geometry and leave the detailed construction of the full interpolating solution unspecified.

 \begin{table}[htbp]
        \centering
        \addtocounter{footnote}{-1}
        \begin{tabular}{c c c c}
            \toprule
            $\theta$ range & $d_{\text{eff}}$ & Scaling of entanglement quantities & Thermal entropy scaling \\
            \midrule
            $\theta < d-1$ & $>1$ & $l^{1-d_{\text{eff}}}$ (decays with $l$) & $T^{d_{\text{eff}}}$ (increases with $T$) \\
            $\theta = d-1$ & $=1$ & $\log l$ or $l^0$ (logarithmic or constant) & $T$ (linear) \\
            $d-1 < \theta < d$ & $<1$ & $l^{1-d_{\text{eff}}}$ (sub-volume law) & $T^{d_{\text{eff}}}$ (decreases with $T$) \\
            $\theta = d$ & $=0$ & $l$ (volume law) & constant ($T^0=1$) \\
            \bottomrule
        \end{tabular}
        %\label{Table1}
        \caption[Thermal entropy and scaling behaviors of entanglement quantities for different $\theta$ ranges. Here we set $z_t=1$. \textcolor{purple}{check the case deff=1, if entanglement and thermal scaling has other forms, add: balanced constant (markov gap, I3), unbalanced log l (multi-entropy)} Note that at $d_{\text{eff}}=1$, some entanglement measures exhibit logarithmic scaling, while others scale as $l^0$, i.e., they exhibit constant scaling behavior.]{Thermal entropy and scaling behaviors of entanglement quantities for different $\theta$ ranges. Here we set $z_t=1$. Note that at $d_{\text{eff}}=1$, some entanglement measures exhibit logarithmic scaling, while others scale as $l^0$, i.e., they exhibit constant scaling behavior\footnotemark.}
        \label{Table1}
    \end{table}
    \footnotetext{This distinction mainly depends on whether the logarithmic divergence of the entanglement qunatity is balanced. For unbalanced measures such as entanglement entropy or multi entropy, which are affected by the UV cutoff, logarithmic scaling appears; for most balanced measures such as $I_3$ or markov gap, constant scaling behavior is observed.}
    % \chen{It should be noted that, although we do not present a general holographic system whose infrared geometry is of the HV type, we have provided the null energy condition (NEC) constraints \eqref{NullEnergyCondition}, which demonstrate that, with appropriate parameter choices and suitable matter fields, a complete geometry interpolating from the IR HV geometry to the UV asymptotically AdS spacetime can indeed be realized. In fact, several concrete constructions along these lines already exist in the literature, to which we refer the interested reader for further details.} {\color{orange} Explicit interpolating solutions of this type have been constructed in both bottom-up Einstein-Maxwell-dilaton models and top-down string-theory settings~\cite{Iizuka_2013,Cremonini_2016}.%1406.5992不太行}
    %When \(d-1 < \theta < d\), the effective dimension becomes a fractional number less than one. In this regime, the entanglement entropy exhibits an unusual sub-volume law scaling as we have demonstrated. The system is neither a gapped phase (where \(d_{\text{eff}}=0\) and all degrees of freedom are frozen) nor a fully extended critical phase (where \(d_{\text{eff}}=1\)), but rather behaves as if it were defined on a fractal set, with its Hausdorff dimension being a fraction. Such behavior is typically encountered in topologically ordered systems or other long-range correlated systems, where the critical physics can be equivalently described by a short-range interacting model on an effective fractal geometry.
   
    After establishing the scaling behavior of entanglement measures in systems whose IR geometry is described by an HV geometry, we now calculate the tripartite information \(I_3\) as a concrete example for further investigation. As pointed out in \cite{ju2024holographicmultipartiteentanglementupper}, \(I_3\) is not, in general, a faithful measure of tripartite entanglement. However, in HEGMEC configurations, where all bipartite information vanishes while only tripartite information remains, \(-I_3\) becomes a meaningful tripartite quantity that isolates the exclusive correlations shared among all three parties. In certain HEGMEC configurations where \(-I_3\) saturates its upper bound, it is entirely quantum in origin. In a generic HEGMEC configuration, by contrast, when \(-I_3\) lies below the classical upper bound, the correlations captured by \(-I_3\) may be either quantum or classical. Nevertheless, \(-I_3\) remains a useful tripartite quantity because it isolates correlations that are exclusively shared among all three parties.
    
    We consider three parallel strips \(A, B, C\) of equal width \(l_{\text{strip}}\), with distances between \(A,B\) and \(B,C\) both set to \(l_{\text{gap}}\), which is to be determined by the HEGMEC condition. By properly tuning the ratio between \(l_{\text{strip}}\) and \(l_{\text{gap}}\), we ensure that any two strips have disconnected entanglement wedges while all three strips together have a connected entanglement wedge. In this configuration, the tripartite mutual information reduces to $I_3(A:B:C) = S_{ABC} - S_{AB} - S_{AC} - S_{BC} + S_A + S_B + S_C = S_{ABC} - S_A - S_B - S_C$. Let \(l_{\text{gap}} = \eta\,l_{\text{strip}}\), where \(\eta = l_{\text{gap}} / l_{\text{strip}}\) is the ratio of the gap width to the strip width. Then the conditions for the {HEGMEC} configuration become
    \begin{equation}
        \nonumber
        \begin{aligned}
            \eta^{1-d_{\text{eff}}} + (\eta+2)^{1-d_{\text{eff}}} &> 2,\\
            2\eta^{1-d_{\text{eff}}} + (2\eta+3)^{1-d_{\text{eff}}} &< 3.
        \end{aligned}
    \end{equation}
    
    \begin{figure}[htbp]
        \centering
        \includegraphics[width=0.55\linewidth]{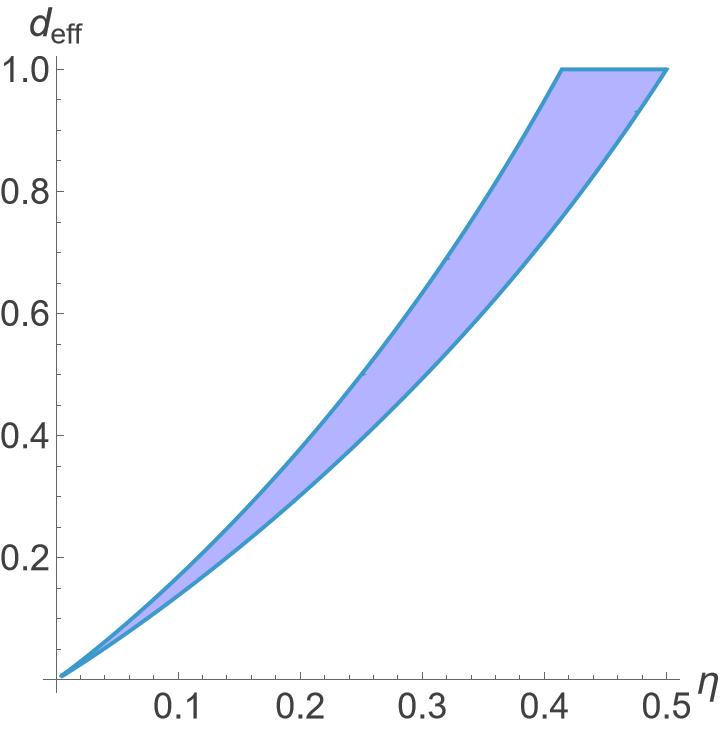}
        \caption{The range of \(\eta\) allowed for the {HEGMEC} configuration under different effective dimensions \(d_{\text{eff}}\), where \(\eta = l_{\text{gap}} / l_{\text{strip}}\) is the ratio of the gap width to the strip width. It is observed that as \(d_{\text{eff}}\) approaches zero, the allowed region for the {HEGMEC} configuration becomes narrower and finally disappears completely at \(d_{\text{eff}}=0\).}
        \label{FigureGhemecRegion}
    \end{figure}

    The region of \(\eta\) satisfying these inequalities is plotted in Fig.\,\ref{FigureGhemecRegion}. Note that $\eta$ depends only on $d_{\text{eff}}$ and not directly on other scaling parameters. We observe that as the effective dimension of the system decreases, the existence window of the {HEGMEC} configuration becomes smaller and smaller, and completely disappears when $d_{\text{eff}}=0$. The underlying reason is that such a configuration requires the Hilbert spaces of the three subsystems to be sufficiently large and the overall volume to be large enough. As $d_{\text{eff}}$ approaches zero, the number of active degrees of freedom at low energies in the boundary field theory decreases, while the correlation length of the system becomes infinitely long. In fact, at low energies the system behaves like a $(0+1)$-dimensional quantum mechanics, i.e., only the time dimension remains. Thus on a time slice, the low-energy limit approximates a single point, making it impossible to define spatially separated subsystems. Therefore, it is expected that the existence window of the {HEGMEC} configuration shrinks to zero as $d_{\text{eff}}$ tends to zero.
    
    Correspondingly, the tripartite mutual information scales as 
    \begin{equation}
        \label{i3eta}
        I_3 \propto \left[ (2\eta+3)^{1-d_{\text{eff}}} + 2\eta^{1-d_{\text{eff}}} - 3 \right] l_{\text{strip}}^{1-d_{\text{eff}}}.
    \end{equation}
    The explicit result in \eqref{i3eta} confirms the general analytic scaling derived in \eqref{Result}. In the present setup, the boundary theory has three spatial dimensions, \(d=3\). For a conventional holographic critical system with an AdS\(_5\) IR geometry, corresponding to \(d_{\mathrm{eff}}=3\), strip-based multipartite entanglement quantities such as \(I_3\) typically decay as \(l_{\mathrm{}}^{-2}\) at large distances. By contrast, when \(0<d_{\mathrm{eff}}<1\), \eqref{i3eta} gives
    \begin{equation}
        \nonumber
        I_3\propto l^{\,1-d_{\mathrm{eff}}},
    \end{equation}
    with a negative coefficient throughout the HEGMEC regime. Therefore, the magnitude \(-I_3\) grows with the overall subsystem size, providing an explicit realization of scale-growing long-range multipartite correlations.

    We further note that the coefficient of this scaling depends explicitly on the shape parameter \(\eta\). This shape dependence indicates that the underlying mechanism is distinct from the long-range entanglement generated by ground-state topological constraints in topologically ordered phases, whose universal topological contribution is independent of the overall scale and, for smooth entangling boundaries, does not depend explicitly on the detailed shape of the configuration.

    At \(d_{\mathrm{eff}}=1\), the system behaves equivalently to a \(\text{CFT}_2\) field theory at long distance scales. \(I_3\) scales as \(l^0\), in agreement with the general prediction in \eqref{Result}. The apparent difference from the logarithmic entanglement entropy of the corresponding effective \(\mathrm{CFT}_2\) originates from their different balance properties. Unbalanced quantities, such as the entanglement entropy and multi-entropy, retain the logarithmic term, whereas balanced quantities, including \(I_3\), the Markov gap, and \(\kappa\), exhibit constant scaling, as summarized in Table~\ref{Table2}.
    \begin{table}[htbp]
        \centering
        \begin{tabular}{c c}
            \toprule
            Entanglement quantities & scaling behavior\\
            \midrule
            Entanglement entropy $S_A$ & $\log\lambda$\\
            Multi entropy $S^{(3)}(A:B:C)$ & $\log\lambda$\\
            Markov gap & $\lambda^0$\\
            $I_3$ & $\lambda^0$\\
            $\kappa$ & $\lambda^0$\\
            \bottomrule
        \end{tabular}
        \caption{We have listed the scaling behaviors of several entanglement measures at $d_{\text{eff}}=1$. It is evident that all unbalanced entanglement measures, which suffer from UV divergences, exhibit logarithmic scaling, whereas the balanced entanglement measures all display constant behavior, precisely as predicted.}
        \label{Table2}
    \end{table}
    
    %In the hyperscaling violating geometry, when $d_{\text{eff}}=0$ the system behaves exactly like a gapped system, with the metric
    %\begin{equation}
    %    ds^2 = -r^2 dt^2 + \frac{dr^2}{r^4} + \sum_{i=1}^d (dx^i)^2,
    %\end{equation}
    %whose IR geometry is homeomorphic to $AdS_2 \times \mathbb{R}^d$. However, due to the nature of HV geometries, regardless of the values of the hyperscaling violation parameters $\theta$ and $z$, for a strip region of width $l_i$ on the boundary, the turning point depth $r_*$ of the corresponding RT surface always scales as $r_* \propto l_i^{-1}$ at large scales. In other words, unlike the phase transition observed in the $xy$ directions of the $AdS_3\times\mathbb{R}^2$ IR geometry — where the RT surface changes from an arch homeomorphic to the boundary strip to two disconnected walls as $l_i$ increases — such a transition does not occur in HV geometries. In summary, when $d_{\text{eff}}=0$, the HV geometry yields the \textit{longest-range entanglement} reminiscent of thermal entropy.

    We next clarify the physical origin of the scale-growing entanglement in the regime \(0<d_{\mathrm{eff}}<1\). As emphasized above, this behavior is conceptually different from the long-range entanglement of a gapped topologically ordered phase. In a topologically ordered state, the long-range entanglement is  protected by topology and is shape independent. The behavior found here instead occurs in a gapless state and remains explicitly dependent on the length scale and the shape of the configurations, which does not possess a topological origin. 
    
    For a conventional scale-invariant IR fixed point in a \(3+1\)-dimensional gapless system, the long-distance multipartite entanglement associated with parallel strips is expected to decay as \(l^{-2}\). Here, \(l\) denotes the finite width of each strip along the \(x\) direction, while the strips are infinitely extended along the two transverse directions \(y\) and \(z\). This scaling follows directly from dimensional analysis. For a fixed-shape multipartite configuration, all strip widths and separations are scaled simultaneously with \(l\), while all dimensionless shape ratios are kept fixed. After the local ultraviolet area-law contributions cancel, an ordinary \(3+1\)-dimensional scale-invariant IR theory contains no remaining length scale other than \(l\). Therefore, any finite multipartite quantity must take the form
    \begin{equation}
        Q_{\mathrm{multi}}(l)\sim\frac{\mathcal{A}_{\perp}}{l^{2}},
        \label{eq:ordinary_multipartite_scaling}
    \end{equation}
    where \(\mathcal{A}_{\perp}\) is the regulated transverse area independent of $l$ and the coefficient may depend on the dimensionless parameters specifying the shape of the multipartite configuration.

    Within the class of local, scale-covariant holographic ground states described by a smooth classical IR geometry, a scale-growing term therefore requires a modification of the IR scaling in the entropy quantities. Hyperscaling violation provides a minimal realization of such a modification. For an isotropic hyperscaling-violating geometry, the relevant scaling is characterized by the effective spatial dimension
    \begin{equation}
        d_{\mathrm{eff}}=d-\theta.
    \end{equation}
    Suppressing the fixed dimensionful normalization associated with the hyperscaling-violating regime, the corresponding finite entanglement contribution scales as
    \begin{equation}
        Q_{\mathrm{multi}}(l)\sim\mathcal{A}_{\perp}l^{1-d_{\mathrm{eff}}},
    \end{equation}
    up to a shape-dependent coefficient and possible cancellations of the leading IR contribution.

    This scaling admits a simple field-theoretic interpretation in terms of effective degrees of freedom. At a length scale \(\xi\), the ground state may be coarse-grained into collective degrees of freedom that are coherent over distances of order \(\xi\). In an IR fixed point with effect spatial dimension \(d_{\mathrm{eff}}\), under a coarse-graining step \(\xi\rightarrow b\xi\), the effective density of low-energy degrees of freedom decreases by \(b^{-d_{\mathrm{eff}}}\), whereas the strip volume associated with this scale increases by \(b\). Their net number therefore changes by \(b^{1-d_{\mathrm{eff}}}\). For \(d_{\mathrm{eff}}>1\), long-wavelength degrees of freedom become progressively less important. At \(d_{\mathrm{eff}}=1\), their number remains of the same order at each scale. For \(0<d_{\mathrm{eff}}<1\), the growth of the spatial region exceeds the reduction produced by coarse graining, leaving an increasing effective number of collective degrees of freedom available at longer scales. This gives the scale-growing multipartite contribution $Q_{\mathrm{multi}}(l)\sim\mathcal{A}_{\perp}l^{1-d_{\mathrm{eff}}}$. At \(d_{\mathrm{eff}}=0\), the effective density no longer decreases with the coarse-graining scale at the level of this scaling description. The number of IR degrees of freedom then grows in proportion to the strip volume, leading to a volume-law multipartite contribution. The effective dimension therefore characterizes how rapidly degrees of freedom are removed from the low-energy sector under renormalization: the smaller \(d_{\mathrm{eff}}\) is, the more degrees of freedom remain available to support correlations over long distances.

    It is important to note that this interpretation specifies the scaling distribution of the IR degrees of freedom, rather than their microscopic identity. Different microscopic theories may produce the same effective value of \(d_{\mathrm{eff}}\) through different strongly coupled sectors, fractionalized degrees of freedom, or other mechanisms encoded in the ultraviolet completion. 

    Note that the limiting case \(d_{\mathrm{eff}}=0\) should not be identified with the holographic semi-local criticality in near horizon AdS$_2$ geometries \cite{Iqbal2012}. Within the finite-\(z\) hyperscaling-violating family, \(d_{\mathrm{eff}}=0\) corresponds to \(\theta=d\). At this point, the entanglement scaling becomes extensive and the thermal entropy density scales as
    \begin{equation}
        s(T)\sim T^{d_{\mathrm{eff}}/z}\sim T^{0}.
    \end{equation}
    Thus, this limit shares with an \(\mathrm{AdS}_{2}\times\mathbb{R}^{d}\) geometry the presence of a finite zero-temperature entropy density. This thermodynamic similarity, however, does not imply that the two IR structures are equivalent.

    For finite \(z\), the hyperscaling-violating geometry remains scale covariant in the spatial directions,
    \begin{equation}
        x^{i}\longrightarrow\lambda x^{i},\qquad t\longrightarrow\lambda^{z}t.
    \end{equation}
    The boundary theory therefore retains spatially extended scale-invariant degrees of freedom, even when \(d_{\mathrm{eff}}=0\). By contrast, semi-local criticality is characterized by nontrivial scaling only in the temporal direction, while the spatial coordinates do not scale,
    \begin{equation}
        x^{i}\longrightarrow x^{i},\qquad t\longrightarrow\lambda t.
    \end{equation}
    Holographically, this behavior is associated with an\(\mathrm{AdS}_{2}\times\mathbb{R}^{d}\) region or, more generally, with a geometry conformal to \(\mathrm{AdS}_{2}\times\mathbb{R}^{d}\). In the standard hyperscaling-violating parametrization, such semi-local geometries arise from a separate double-scaling limit in which \(z\rightarrow\infty\) and \(\theta\rightarrow-\infty\), with an appropriate ratio held fixed, rather than from the finite-\(z\) condition \(\theta=d\).

    Therefore, \(d_{\mathrm{eff}}=0\) should be understood as the volume-law endpoint of the spatially scale-covariant hyperscaling-violating family. It may reproduce some thermodynamic features of a semi-local phase, but it does not by itself imply an \(\mathrm{AdS}_{2}\) IR symmetry, a finite spatial correlation length, or purely temporal criticality.

    Finally, it is worth noting that although all entanglement measures constructed from minimal surfaces share the same scaling behavior for a given IR geometry, they do not necessarily probe the same entanglement structure. For example, $\kappa$ is sensitive to GHZ-type correlations but blind to W-type ones, while the Markov gap exhibits the opposite sensitivity \cite{https://doi.org/10.48550/arxiv.2411.11961,Bao2026}. This raises the question of why they nevertheless scale identically. One possibility is that the holographic IR geometry encodes a universal multipartite entanglement pattern, and different measures merely project onto different facets of this same underlying structure. Another possibility is that, even if the measures probe genuinely distinct structures, the holographic dynamics somehow ensures that all such structures contribute at comparable levels in the IR region, so that no particular type of entanglement dominates over others in the long-range scaling. Our present analysis does not distinguish between these two scenarios, and it would be interesting to further investigate which interpretation is realized in specific holographic models.

%%%%%%%%%%%%%%%%%%%%%%%%%
\section{Anisotropic Long-Range Entanglement from an AdS\(_3\times \mathbb{R}^2\) IR Geometry}
    \label{Section4}
    In the previous section, we studied isotropic HV IR geometries, and found  enhanced long-range multipartite entanglement in a specific range of the IR effective dimension $d_{\text{eff}}$, so that a single $d_{\text{eff}}$ governs the leading large-distance behavior in all spatial directions. It is therefore natural to ask whether IR dimensional reduction can be direction-dependent, with more interesting and richer long-range entanglement behavior. This motivates us to go beyond isotropic HV geometries and  further investigate anisotropic hyperscaling violating geometries. 

    In the anisotropic HV geometry \eqref{HyperscalingViolatingGeometry}, we introduce a set of parameters including $z_t$, $z_i$, and $\theta$. To ensure that the metric components $g_{ii}, g_{tt}$ at the horizon are not divergent, we always impose $z_i\ge 0$ and $z_t>0$. In the limit $\theta=0$ and $z_t=z_i=1$, the geometry reduces to the standard AdS case. In isotropic systems, the quantity $d-\theta$ characterizes the low-energy effective dimension of the system; in the anisotropic case, $(\sum_{i=1}^d z_i - \theta)/z_i$ plays an analogous role in corresponding directions in determining the behavior of entanglement quantities. We impose a restriction on the parameter $\theta$ as well, requiring $0\le\theta\le\sum_{i=1}^d z_i$, analogous to the requirement $0\le\theta\le d$ in the isotropic case. Besides, all the parameters need to satisfy the NEC in \eqref{NullEnergyCondition}.

    For a system whose IR geometry is given by the isotropic HV geometry, the scaling behavior of the large-$l$ (i.e. $l$-dependent) part of the entanglement measures for strip configurations can be read off directly from the preceding discussion, as summarised in Table~\ref{Table3}. {When the parameters satisfy $0<z_i = \sum_{j=1}^d z_j-\theta$, the entanglement measures exhibit logarithmic or constant scaling behavior; in the regime $0<\sum_{j=1}^d z_j-\theta<z_i$, they can further display sub-volume-law scaling.} Within our specified parameter range, the longest-range entanglement behavior of these measures is also volume-law scaling. We additionally assume that all $z_i$ are nonzero; the special case where some $z_i=0$ will be treated separately below.
    \begin{table}[htbp]
        \centering
        \begin{tabular}{c c}
            \toprule
            parameter range & Scaling of entanglement quantities \\
            \midrule
            $0<z_i < \sum_{j=1}^d z_j-\theta$ & $l^{1-\frac{\sum_{j=1}^d z_j-\theta}{z_i}}$ (decays with $l$) \\
            $0<z_i = \sum_{j=1}^d z_j-\theta$ & $\log l$ or $l^0$ (logarithmic or constant) \\
            $0<\sum_{j=1}^d z_j-\theta<z_i$ & $l^{1-\frac{\sum_{j=1}^d z_j-\theta}{z_i}}$ (sub-volume law) \\
            $0=\sum_{j=1}^d z_j-\theta<z_i$ & $l^{1}$ (volume law) \\
            \bottomrule
        \end{tabular}
        \caption{{The large-scale scaling behavior of the $l$-dependent part of entanglement measures for a strip oriented along the $x^i$ direction in different parameter regimes, with all $z_i$ being nonzero. $(\sum_{j=1}^d z_j - \theta)/z_i$ plays a role analogous to $d_{\text{eff}}$ in the isotropic HV geometry. When $\sum_{j=1}^d z_j - \theta = z_i$, the behavior may be either constant or logarithmic, depending on whether the entanglement qunatity is balanced.}}
        \label{Table3}
    \end{table}

    In previous works, such as those on holographic Weyl semimetals \cite{Landsteiner2016} and holographic nodal line semimetals \cite{Liu2021}, the IR geometry is precisely an anisotropic HV geometry with $\theta=0$ and $z_i>0$ for all directions. We numerically computed various entanglement measures along different directions, including $\kappa$, the Markov gap, and the multi-Markov gap, among others \cite{chen2026,chen2026detectingtopologicaltransitionsanisotropy}. The long-range scaling behaviors of these measures indeed follow those summarized in Table~\ref{Table3} and exhibit clear anisotropy. Moreover, these holographic systems with HV IR geometry provide a complete flow profile from the IR HV geometry to the UV asymptotically AdS geometry. While explicit interpolating solutions from anisotropic HV IR geometries to asymptotically AdS$_5$ ultraviolet geometries have not been constructed in this work, we believe that with suitable matter fields such full flows can indeed be realized. Explicit constructions in previous literature could be found in \cite{Giataganas_2026,Mateos_2011,Jeong_2018}.

    When $z_i=0$, the exponent $(\sum_{j=1}^d z_j-\theta)/z_i$ becomes singular, and hence the long-range scaling behavior summarized in Table~\ref{Table3} no longer applies. We therefore now turn to examine explicitly the long-range scaling behavior of entanglement measures in the case where one of the directions has $z_i=0$. For an anisotropic HV geometry, the scaling components combination $1-\alpha_i+\alpha_r = z_i$. For $z_i=0$, {as discussed in Section~\ref{Section2}, the RT surface of a strip oriented along this direction undergoes a phase transition: when the strip width exceeds a critical value, the connected arch-shaped surface turns into two disconnected walls. This implies that beyond a characteristic length scale, the system loses correlation along this direction, i.e., the system is gapped in that direction. For the case of spatial dimension $d=3$ and for simplicity we set $\theta=0$, we can set $z_i=0$ in one, two, or all directions. These three different cases correspond respectively to IR geometries of AdS$_4\times\mathbb{R}^1$, AdS$_3\times\mathbb{R}^2$, and AdS$_2\times\mathbb{R}^3$. The AdS$_4\times\mathbb{R}^1$ case has no long-range entanglement, while the AdS$_2\times\mathbb{R}^3$ case has all three directions gapped with no gapless long range entanglement.  Hence, we take AdS$_3\times\mathbb{R}^2$ as a concrete example to examine its long range multipartite entanglement properties in detail.
   
    Without loss of generality, we investigate the following holographic model {as a minimal realization}, which allows $\text{AdS}_3\times\mathbb{R}^2$ as an IR geometry \cite{DHoker2009, Sun:2016gpy}
    \begin{equation}
        \begin{split}
            S &= \frac{1}{16\pi G}\int d^5x\sqrt{-g}\Bigl(  R + F^{ab}F_{ab} - \frac{12}{L^2} - \alpha \epsilon^{abcde} A_a F_{bc} F_{de} \Bigr),
        \end{split}
    \end{equation}
    where $A$ is a vector field, $F$ is the corresponding field strength, $G$ is the gravity constant and $R$ is Ricci scalar field. $L$ is set as 1. The Chern--Simons term proportional to $\alpha$ was included in \cite{DHoker2009, Sun:2016gpy} for purposes specific to the original construction and makes no contribution to the dynamics here. 

    To realize an \(\mathrm{AdS}_3 \times \mathbb{R}^2\) geometry in the    IR, with the \(\mathbb{R}^2\) factor spanning the transverse \(x\)-\(y\) plane, we turn on a homogeneous magnetic field directed along the \(z\)-axis, $F = B dx \wedge dy$. A metric ansatz consistent with the translational symmetry and the residual rotational symmetry in the \(x\)-\(y\) plane is
    \begin{equation}
        ds^2=-u(r)\,dt^2+ f(r)\left(dx^2+dy^2\right)+ u(r)\,dz^2+ \frac{dr^2}{u(r)},
    \end{equation}
    where \(u(r)\) and \(f(r)\) are real functions of the holographic radial coordinate \(r\). Substituting this ansatz and the magnetic-field configuration into the bulk equations of motion, the independent equations governing \(u(r)\) and \(f(r)\) can be written as
    \begin{equation}
        \begin{aligned}
            \frac{u''}{u}-\frac{f''}{f}+\frac{u'^2}{2u^2}-\frac{u'f'}{2uf}&=\frac{4B^2}{uf^2},\\
            \frac{u''}{u}+\frac{u'f'}{uf}+\frac{u'^2}{2u^2}-\frac{8}{u}&=\frac{4B^2}{3uf^2}.
        \end{aligned}
    \end{equation}

    The equations of motion admit the IR fixed-point solution
    \begin{equation}
        u(r)=3r^2,\qquad f(r)=\frac{B}{\sqrt{3}},
    \end{equation}
    which can also be extended as an exact solution over the entire radial domain. However, this solution is not asymptotically \(\mathrm{AdS}_5\), since \(f(r)\) approaches a constant rather than scaling as \(r^2\) near the UV boundary. To construct a geometry that interpolates between \(\mathrm{AdS}_3\times\mathbb{R}^2\) in the IR and \(\mathrm{AdS}_5\) in the UV, we impose the asymptotically AdS boundary conditions
    \begin{equation}
        \lim_{r\to\infty}\frac{u(r)}{r^2}=\lim_{r\to\infty}\frac{f(r)}{r^2}=1.
    \end{equation}

    Starting from the IR solution, we introduce a small deformation \(\delta r\) in \(f(r)\), with the corresponding correction to \(u(r)\) determined by the equations of motion. The parameter \(\delta\) then serves as a shooting parameter. By tuning \(\delta\) such that both \(u(r)\) and \(f(r)\) approach \(r^2\) at large \(r\), we obtain a complete bulk solution that interpolates smoothly from an \(\mathrm{AdS}_3\times\mathbb{R}^2\) geometry in the deep IR to an asymptotically \(\mathrm{AdS}_5\) geometry in the UV.

    Under the given conditions, from the preceding discussion we know that the system is gapped in both the $x$ and $y$ directions, while in the $z$ direction it behaves effectively as a CFT$_2$. To further investigate its properties, we shall compute the $c$-function and the tripartite mutual information $I_3$ for strips oriented along the $z$ and $x$ directions as representative quantities.

    We first calculate the $c$-function for the system, which reflects the change in degrees of freedom during the coarse-graining process of the renormalization group flow from UV to IR. For a strip region of width $l_i$ along the $x^i$ direction, the corresponding $c$-function is $c_i = \frac{l_i^3}{2}\frac{dS_i}{dl_i}$ \cite{Myers2012,Baggioli2021,Baggioli2023}, where $S_i$ is the entanglement entropy of the strip region. We computed the $c$-functions $c_x$ and $c_z$ for the $x$ and $z$ directions in this system and fitted their large-scale power-law behavior. We find $c_z \propto l_z^2$ while $c_x$ is strictly zero at large $l_x$. This indicates that the system indeed opens a long-range correlation channel in the $z$ direction, while the degrees of freedom in the $xy$ plane are almost completely frozen.

    To further characterize the behavior in different directions, we computed the tripartite mutual information $I_3$ in the HEGMEC configuration. We chose three parallel strips in the $x$ direction and in the $z$ direction, respectively, and tuned the ratios of strip widths and separations such that any two strips have disconnected entanglement wedges while the three strips together have a connected entanglement wedge. In the $z$ direction, $I_3$ is strictly a constant, consistent with the tripartite mutual information of the HEGMEC configuration in a CFT$_2$. In the $x$ direction, due to the phase transition of the RT surface at large $l_x$, all entanglement follows a strict area law, meaning that no HEGMEC configuration exists. Therefore, we conclude that the physics of this system corresponds to all long-range entanglement in the $x$ and $y$ directions being blocked, while long-range CFT$_2$-type entanglement emerges in the $z$ direction.

    \begin{figure}[htbp]
        \centering
        \includegraphics[width=0.48\linewidth]{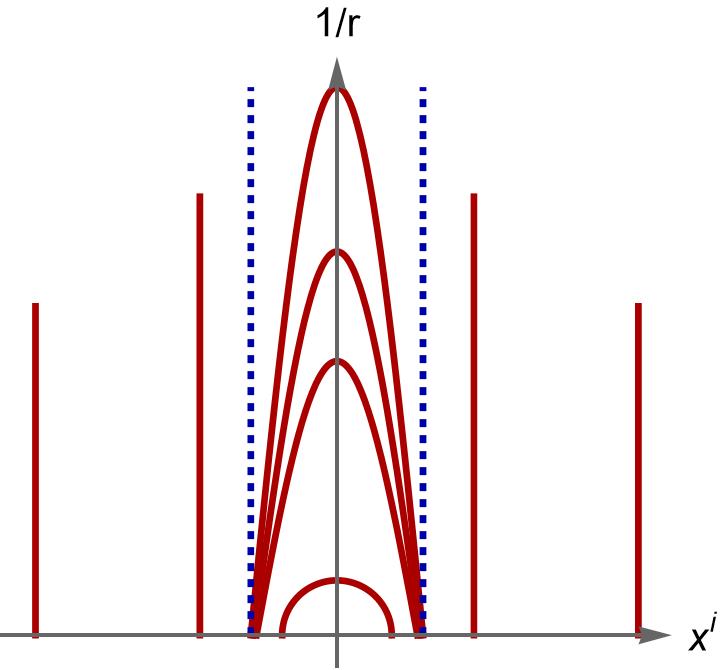}
        \caption{RT surface for a strip of width $l_i$ along a gapped $x^i$ direction. Beyond a critical width, indicated by the blue dashed lines, the connected surface is replaced by two disconnected walls. The entanglement entropy then becomes independent of $l_i$ and obeys a pure area law, signaling a finite correlation length and suppressed low-energy excitations along this direction.}         
\label{FigurePhaseTransition}
\end{figure}
 
    In brief, in such a system, the application of a magnetic field along the $z$-direction leads to Landau quantization of the charged degrees of freedom in the transverse $x$--$y$ plane. As a result, in the deep IR regime, transverse propagation is frozen to a finite correlation length, leaving no IR channel for long-range propagation in the transverse directions. Consequently, the connected branch of the RT surface becomes unstable beyond a finite critical length as shown in Fig.\,\ref{FigurePhaseTransition}, and the entanglement quantities enter a pure area-law phase. Meanwhile, low-energy propagation persists along the magnetic-field direction. This sector is encoded by an $\mathrm{AdS}_3$ factor in the bulk, which gives rise to a family of effective $(1+1)$-dimensional gapless channels along the $z$-direction. Therefore, the entanglement entropy in the $z$-direction exhibits a logarithmic scaling, and the corresponding tripartite mutual information $I_3$ approaches a constant.

    Closely related anisotropic low-energy behavior occurs in quasi-one-dimensional quantum materials, which may be viewed, over an intermediate range of energies and length scales, as one-dimensional chains. In the idealized decoupled-chain limit, transverse correlations are short-ranged and the corresponding long-distance entanglement is boundary dominated, whereas gapless critical correlations can persist along each chain. The spin-chain cuprate $\text{Sr}_2\text{CuO}_3$ provides an example: above the low-energy crossover scale set by weak interchain coupling, it is well described by an $S=1/2$ Heisenberg chain with gapless spinon excitations, whose continuum limit is a $(1+1)$-dimensional conformal field theory \cite{Sologubenko2000}. The Kondo-lattice compound $\text{CeCo}_2\text{Ga}_8$ is likewise strongly anisotropic and quasi-one-dimensional, and therefore provides a qualitative analogue of direction-selective low-energy dynamics \cite{Cheng2019}, though its IR critical theory has not been established to be a simple $\text{CFT}_2$. These materials should therefore be regarded as condensed-matter analogues of the anisotropic IR structure that exhibit the same long-range entanglement behavior in gapless systems.
%%%%%%%%%%%%%%%%%%%%%%%%%
\section{Conclusion and discussions}
    \label{Section5}
    In this work, we studied long-distance multipartite entanglement in holographic gapless systems. We showed that, for static, translationally invariant, asymptotically AdS geometries, the leading large-distance power of any nonvanishing multipartite entanglement quantity constructed from a finite network of minimal surfaces is determined by the IR geometry. For a general set of isotropic hyperscaling-violating IR geometries, this power is controlled by the IR effective dimension \(d_{\mathrm{eff}}=d-\theta\). The long-distance contribution decays for \(d_{\mathrm{eff}}>1\), becomes logarithmic at \(d_{\mathrm{eff}}=1\), and grows with the subsystem size for \(0<d_{\mathrm{eff}}<1\), reaching a volume law at \(d_{\mathrm{eff}}=0\). Thus, suitable IR geometries can support scale-growing long-distance multipartite entanglement even in a gapless state. The resulting scale and shape-dependent long-range entanglement is distinct from the scale or shape-independent contribution associated with topological order.

    We also considered anisotropic HV IR geometries. In the magnetic-brane geometry flowing from \(\mathrm{AdS}_{5}\) to \(\mathrm{AdS}_{3}\times\mathbb{R}^{2}\), long-range entanglement survives along the effective \((1+1)\)-dimensional direction, whereas the transverse directions exhibit only short-range, area-law entanglement beyond a critical strip width.  Taken together, our results demonstrate that holographic gapless systems can support parametrically enhanced long-range multipartite entanglement. Depending on the IR geometry, this entanglement may decay, exhibit marginal scaling, grow subextensively, or reach a volume law, while anisotropic IR geometries can further restrict it to selected spatial directions. Long-range multipartite entanglement therefore constitutes a broad and geometrically organized feature of holographic gapless systems, rather than an exception tied only to topological order. It also provides a nonlocal diagnostic of the IR universality class, including its effective dimension and anisotropic dimensional-reduction structure.
    
    Several natural directions follow from our analysis. First, our derivation relies on strip configurations with translational symmetry. Extending the analysis to more general shapes of entangling regions would reveal richer underlying multipartite entanglement structures. Second, it would be interesting to investigate whether the long range entanglement behaviors are accompanied by characteristic signatures in charge and heat transport, thereby establishing a closer connection between quantum information and measurable response functions. Third, it remains unclear whether the universal scaling of all entanglement measures originates from a single underlying entanglement structure projected onto different quantities, or from a holographic mechanism that equally amplifies distinct structures in the IR. It will be an important open question to distinguish these possibilities. Finally, extending the present framework to time-dependent settings, such as quantum quenches or driven holographic states, would reveal whether the IR scaling structure survives beyond equilibrium and how it is modified by nonequilibrium dynamics.

%%%%%%%%%%%%%%%%%%%%%%%%%
\nocite{*}
%%%%%%%%%%%%%%%%%%%%%%%%%

%%%%%%%%%%%%%%%%%%%%%%%%%%%%%%
\subsection*{Acknowledgments}
%%%%%%%%%%%%%%%%%%%%%%%%%%%%%%
We thank Yan Liu, Wen-Bin Pan, Bo-Yu Xu and Yang Zhao for helpful discussions. XXJ would like to thank the support of the
Shuimu Tsinghua Scholar Program of Tsinghua University. This work was supported by the National Natural Science Foundation of China (Grant Nos. 12575068, 12405078). 
%%%%%%%%%%%%%%%%%%%%%%%%%%%
%\include{Appendices}
%%%%%%%%%%%%%%%%%%%%%%%%%%%

%
\bibliographystyle{elsarticle-num}
\bibliography{biblio}

\end{document}